\documentclass[aps,prl,onecolumn,groupedaddress,showpacs]{revtex4}
\usepackage{revsymb}
\usepackage{epsfig,graphics}
\usepackage{amsfonts}
\def\R{{\mathbb R}}
\begin{document}
\title{Fluid/solid transition in a hard-core system}
\author{Lewis Bowen}
\email[Electronic address: ]{lpbowen@indiana.edu}
\affiliation{Department of Mathematics, Indiana University, Bloomington,
IN}
\author{Russell Lyons}
\email[Electronic address: ]{rdlyons@indiana.edu}
\affiliation{Department of Mathematics, Indiana University, Bloomington,
IN}
\author{Charles Radin}
\email[Electronic address: ]{radin@math.utexas.edu}
\affiliation{Department of Mathematics, University of Texas, Austin, TX 78712}
\author{Peter Winkler}
\email[Electronic address: ]{peter.winkler@dartmouth.edu}
\affiliation{Department of Mathematics, Dartmouth College, Hanover, NH}

\date{\today}
\begin{abstract}
We prove that a system of particles in the plane, interacting only with a certain
hard-core constraint, undergoes a fluid/solid phase transition.
\end{abstract}

\maketitle

\section{I. Introduction}
It is generally believed that the classical
statistical mechanics of particles in $\R^3$ interacting through a
strong, short-range repulsion and weak, short-range attraction would
exhibit a solid/fluid phase transition. It has not been
possible to control this analytically, and this has been an important open
problem for many years \cite{Br}, \cite{U},\cite{S},\cite{Ra},\cite{A}.
There were proofs long ago within lattice gas models of
order/disorder transitions \cite{LY}, \cite{OP}, as well as convincing molecular
dynamics and Monte Carlo simulations showing a transition in
continuum models (particles in $\R^3$; see \cite{Ba} for a review).
In particular, simulations indicate such a transition even
for the conceptually simple hard-sphere model in which the interaction
is just a hard core, and even in 2 dimensions though the ordered phase
may not be crystalline in 2 dimensions \cite{J}.

Two notable analytic proofs of a phase transition in a system of
particles moving in a continuum rather than a lattice were those by
Ruelle \cite{Ru} and the recent paper by Mazel et al.\ \cite{MLP}. A weakness of
the first is that it is more concerned with breaking the discrete
symmetry between particle species than the spatial symmetry. A weakness of the latter is that it
uses a long-range attraction and therefore, like the simpler
Curie-Weiss model, it is more relevant to the gas/liquid transition \cite{MLP}.

The melting transition concerns an ordered structure breaking apart.
This is often modeled as a competition between the influences of energy $E$ and
of entropy $S$ on the distribution over configurations which minimizes
the (Helmholtz) free energy $F:=E-TS$ \cite{SS}. But this competition cannot be the
mechanism for a system with only hard-core forces since the energy is then
just kinetic and can be integrated out; for such
a model, the mechanism must be purely geometric, a competition
between random and ordered configurations as the dominant
contribution to the entropy.

One attempt to give a geometric mechanism for an order/disorder
transition is through ``orientational order'' in the two dimensional
hard-disk model, based on the supposed difficulty or cost of rotating
a pair of neighboring disks past other neighbors \cite{J}. Assuming this
mechanism is significant, a geometric proof of a transition should be
simpler if the relative orientation between neighboring pairs were
constrained, as this would make it easier to detect and quantify long range
order.

In this paper we simply change the shape of the disks in the
hard-disk model, introducing three well-defined levels of constraint
for the relative orientation between neighboring pairs and allowing a
proof to go through. We call the shape a
molecule, and it is basically a solid unit hexagon, but with a fringe of
projections and holes on its edges (see Figure 1) which can accept the 
holes and projections of neighboring molecules in one
of two modes, ``tightly linked'' and ``loosely linked'', the latter
when the projection is less than half way into the neighboring
molecule. (See Figure 2.) 

\begin{figure*}
\epsfig{file=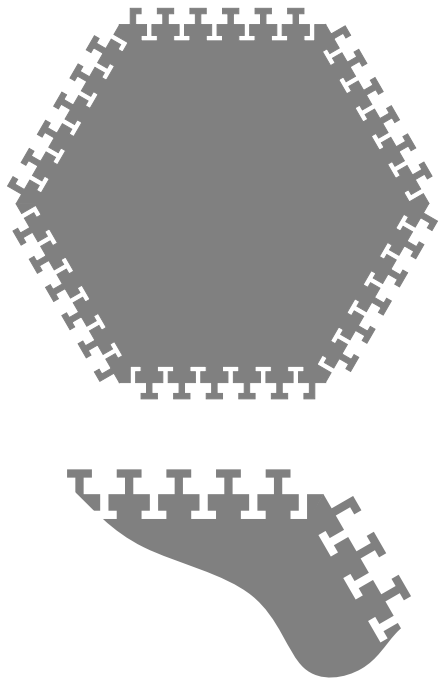,height=5truein}
\caption{The ``zipper'' molecule, including blow-up of a corner.}
\end{figure*}

\begin{figure*}
\epsfig{file=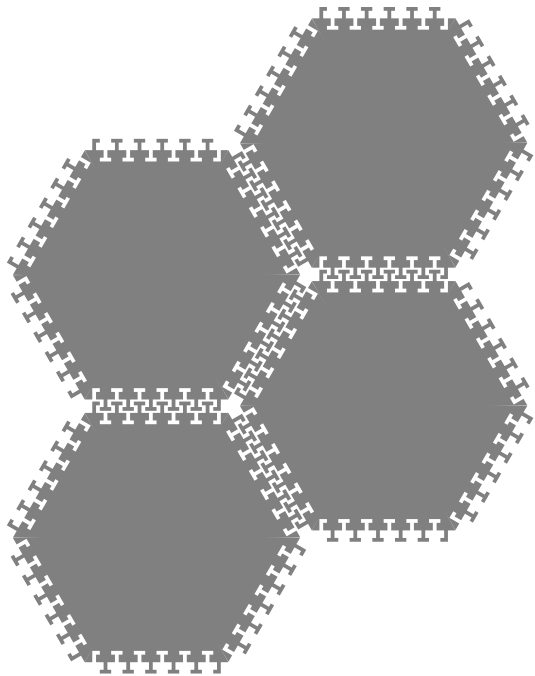,height=3.5truein}
\caption{Loose linked molecules.}
\end{figure*}

We assume the molecule has area 1, so
that number density coincides with packing density. 
A molecule $F$ will be called
``fully linked'' if all its projections and holes are linked
(either tightly or loosely) 
to neighboring molecules. Molecules not linked to other molecules are
called ``free'', and molecules which are neither fully linked nor free
are called ``partially linked''. Each fully linked molecule is
contained in a unique maximal connected set of such molecules
(connected by links). For each packing we decompose the container into the
Voronoi cells of the molecules and concentrate on the
connected components of cells associated to fully linked molecules.

It is easy to see that there exist $1>d_1>d_2>d_3>0$ such that
the following hold: a large region of fully linked molecules in which
all links are loose has density $d\in
[d_2,d_1]$; and the density $d$ of any large collection of
free molecules satisfies $d\le d_3$. 
We assume $d_1$ is
the smallest possible upper limit for the corresponding interval.

We are concerned with the infinite-volume canonical ensemble defined
as the limit of canonical distributions in an expanding sequence of finite boxes
with periodic boundary conditions.
In the finite box, our canonical probability distribution (restricted
to physical space variables; we integrate out the momentum variables,
as usual) is uniform on the set of all arrangements of molecules with
fixed density $d'$ (where $d'\to d$ in the limit). Among those
packings of a given box with fixed density, sets of 
configurations in which the molecules
form fewer tightly-linked components live in a lower-dimensional 
subspace than those with more, and by any
natural approach would have to be accorded zero relative probability;
this would be the case, for instance, if the model were considered the limit of models in
which neighboring molecules could not fit together so perfectly, or of models
with softened core. Thus, a key feature of our model is that the
canonical distribution for a finite box at given density is
supported on those arrangements of molecules with the largest number
of degrees of freedom.

\section{II. Results.} 
Our main result is the following.

\medskip

\noindent {\bf Theorem}. Let $P(d)$ be the probability, given by the
infinite volume canonical ensemble at density $d$,
that the origin (or any given point) is inside an infinite fully-linked cluster. Then there is some $d_4>0$ such that $P(d) = 0$ 
if $d\in (0,d_4)$, while $P(d)>0$ for $d\in (d_1,1)$.

\medskip

\noindent {\bf Corollary}. The model exhibits a fluid/solid phase transition.

\medskip

\noindent {\it Outline of proof of Theorem.}
We begin with high densities.
We shall work directly with invariant measures on configurations in the
whole plane, rather than in finite boxes.
Our molecules have the property that two of them that are
not tightly linked have centers at distance at least $2r+2\rho$, where $r$
is the inradius and $2\rho$ is the fringe height.
At the molecules' densest packing without tight links, the disks of radius
$r+\rho$ concentric with the molecules form a hexagonal close packing,
whence there is a unique invariant measure $\lambda_1$ on packings of
molecules whose density is the highest possible out of those without tight links.  This measure $\lambda_1$ has a number of important
optimization properties, the detailed proofs of which will be published
elsewhere.  First, by choosing $\rho$ to be sufficiently small, we can assume that the area of the
Voronoi cell of the center of a molecule without tight links is at least
$1/d_1$, with equality if and only if its neighbors are arranged as in $\lambda_1$.
To express the other properties, we need some more notation.

If $P$ is a packing with a molecule $m$ containing the origin, let $j(P)$
be the number of molecules in the tightly linked connected cluster
containing $m$ 
and $f(P) := 3/j(P)$, which is
the number of degrees of freedom per molecule for these molecules. Let $j(P) := 
f(P):=0$ if the origin is not contained in a molecule.
Let $\lambda_0$ be the unique invariant measure on tilings by the molecules.
Let $\nu$ be an invariant measure giving average density $d > d_1$.
If we put $s := (1-d)/(1-d_1)\in [0, 1]$ and $\mu := s \lambda_1 +
(1-s) \lambda_0$, then the average density with
respect to $\mu$ equals that with respect to $\nu$.
If $\rho$ is sufficiently small, then 
$\int f\,d\mu \ge \int f\,d\nu$, with equality only if $j(P) \in \{0,
1, \infty\}$ for $\nu$-almost every $P$,
and only if, when $j(P) = 1$, the area of the Voronoi cell of the molecule
containing the origin equals $1/d_1$.
It follows that if $\int f\,d\mu = \int f\,d\nu$,
then $\nu = \mu$.

This implies that the thermodynamic limit distribution is $\mu$.
In particular, $P(d) = 1$.

For low density we can compare a block of
fully linked molecules with a geometrically similar collection of
free molecules in which each has twice the room to move as do the
linked molecules. (This argument ignores the presence of molecules
near the linked molecules; this is permissible at small enough $d_4$
for this crude estimate.) Therefore the probability of a block
of $M$ fully linked molecules is less than $(1/3)^{M}$, which goes
to 0 as $M\to \infty$. (We are simply showing that at low density
the canonical ensemble looks like a gas of independent molecules.)
This proves the desired result for low density, and thus the theorem. $\Box$

\medskip

\noindent {\it Proof of Corollary.} We have shown that at
density $d$ the probability that the origin is near an infinite fully-linked block
of molecules is zero for $d\in (0,d_4)$, while it is positive for $d\in (d_1,1)$. 
This implies that $P(d)$ is not analytic in $d$, which we take
as the hallmark of the transition. $\Box$

\medskip

By analogy with the simulations on hard-sphere and hard-disk models [7]
we expect that the interval of $d$ in which $P(d)>0$ continues below $d_1$, which we would
interpret as implying that the 
melting transition in this model occurs at density below
$d_1$, but we do not know how to prove this.

\begin{acknowledgments}
The authors thank the Banff
International Research Station for support at a workshop where we
began the above research. This paper was supported by the NSF under grants numbered
DMS-0406017 and DMS-0352999.
\end{acknowledgments}

\bibliography{prl11}

\begin{thebibliography}{10}
\expandafter\ifx\csname natexlab\endcsname\relax\def\natexlab#1{#1}\fi
\expandafter\ifx\csname bibnamefont\endcsname\relax
  \def\bibnamefont#1{#1}\fi
\expandafter\ifx\csname bibfnamefont\endcsname\relax
  \def\bibfnamefont#1{#1}\fi
\expandafter\ifx\csname citenamefont\endcsname\relax
  \def\citenamefont#1{#1}\fi
\expandafter\ifx\csname url\endcsname\relax
  \def\url#1{\texttt{#1}}\fi
\expandafter\ifx\csname urlprefix\endcsname\relax\def\urlprefix{URL }\fi
\providecommand{\bibinfo}[2]{#2}
\providecommand{\eprint}[2][]{\url{#2}}

\bibitem[{\citenamefont{Brush}(1983)}]{Br}
\bibinfo{author}{\bibfnamefont{S.~G.} \bibnamefont{Brush}},
  \emph{\bibinfo{title}{Statistical Physics and the Atomic Theory of Matter,
from Boyle and Newton to Landau and Onsager}}
  (\bibinfo{publisher}{Princeton University Press}, \bibinfo{address}{Princeton},
  \bibinfo{year}{1983})
\bibinfo{pages}{277}.

\bibitem[{\citenamefont{Uhlenbeck}(1968)}]{U}
\bibinfo{author}{\bibfnamefont{G.~E.} \bibnamefont{Uhlenbeck}},
  \emph{\bibinfo{title}{Fundamental Problems in Statistical Mechanics II}}
  (\bibinfo{publisher}{Wiley}, \bibinfo{address}{New York},
  \bibinfo{year}{1968})
\bibinfo{pages}{16-17}.

\bibitem[{\citenamefont{Simon}(1984)}]{S}
\bibinfo{author}{\bibfnamefont{B.} \bibnamefont{Simon}},
  \emph{\bibinfo{title}{Perspectives in Mathematics: Anniversary of
Oberwolfach 1984}}
  (\bibinfo{publisher}{Birkhauser Verlag}, \bibinfo{address}{Basel},
  \bibinfo{year}{1984})
\bibinfo{pages}{442}.

\bibitem[{\citenamefont{C. Radin,}(1987)}]{Ra}
\bibinfo{author}{\bibfnamefont{C.} \bibnamefont{Radin}},
\bibinfo{journal}{Int.\ J.\ Mod.\ Phys. B}
  \textbf{\bibinfo{volume}{1}} \bibinfo{pages}{1157-1191}
  (\bibinfo{year}{1987}).

\bibitem[{\citenamefont{Anderson}(1984)}]{A}
\bibinfo{author}{\bibfnamefont{P.~W.} \bibnamefont{Anderson}},
  \emph{\bibinfo{title}{Basic Notions of Condensed Matter Physics}}
  (\bibinfo{publisher}{Benjamin/Cummings}, \bibinfo{address}{Menlo Park},
  \bibinfo{year}{1984}).

\bibitem[{\citenamefont{T.D. Lee and C.N. Yang,}(1952)}]{LY}
\bibinfo{author}{\bibfnamefont{T.~D.} \bibnamefont{Lee}},
\bibinfo{author}{\bibfnamefont{C.~N.} \bibnamefont{Yang}},
\bibinfo{journal}{Phys. Rev.}
  \textbf{\bibinfo{volume}{87}} \bibinfo{pages}{410-419}
  (\bibinfo{year}{1952}).

\bibitem[{\citenamefont{O.J. Heilmann and E. Praestgaard,}(1952)}]{OP}
\bibinfo{author}{\bibfnamefont{O.~J.} \bibnamefont{Heilmann}},
\bibinfo{author}{\bibfnamefont{E.} \bibnamefont{Praestgaard}},
\bibinfo{journal}{J. Stat. Phys.}
  \textbf{\bibinfo{volume}{9}} \bibinfo{pages}{23-44}
  (\bibinfo{year}{1973}).

\bibitem[{\citenamefont{Barker}(1963)}]{Ba}
\bibinfo{author}{\bibfnamefont{J.~A.} \bibnamefont{Barker}},
  \emph{\bibinfo{title}{Lattice Theories of the Liquid State}}
  (\bibinfo{publisher}{Macmillan}, \bibinfo{address}{New York},
  \bibinfo{year}{1963}).

\bibitem[{\citenamefont{A. Jaster,}(1999)}]{J}
\bibinfo{author}{\bibfnamefont{A.} \bibnamefont{Jaster}},
\bibinfo{journal}{Phys. Rev. E}
  \textbf{\bibinfo{volume}{59}} \bibinfo{pages}{2594-2602}
  (\bibinfo{year}{1999}).

\bibitem[{\citenamefont{D. Ruelle,}(1971)}]{Ru}
\bibinfo{author}{\bibfnamefont{D.} \bibnamefont{Ruelle}},
\bibinfo{journal}{Phys. Rev. Lett.}
  \textbf{\bibinfo{volume}{27}} \bibinfo{pages}{1040-1041}
  (\bibinfo{year}{1971}).

\bibitem[{\citenamefont{A. Mazel, J. Lebowitz and E. Presutti,}(1998)}]{MLP}
\bibinfo{author}{\bibfnamefont{A.} \bibnamefont{Mazel}},
\bibinfo{author}{\bibfnamefont{J.} \bibnamefont{Lebowitz}},
\bibinfo{author}{\bibfnamefont{E.} \bibnamefont{Presutti}},
\bibinfo{journal}{Phys. Rev. Lett.}
  \textbf{\bibinfo{volume}{80}} \bibinfo{pages}{4701-4704}
  (\bibinfo{year}{1998}).

\bibitem[{\citenamefont{B. Simon and A. Sokal,}(1981)}]{SS}
\bibinfo{author}{\bibfnamefont{B.} \bibnamefont{Simon}},
\bibinfo{author}{\bibfnamefont{A.} \bibnamefont{Sokal}},
\bibinfo{journal}{J. Statist. Phys.}
  \textbf{\bibinfo{volume}{25}} \bibinfo{pages}{679-694}
  (\bibinfo{year}{1981}).

\end{thebibliography}

\end{document}